%% file: main.tex
\documentclass[a4paper,twocolumn,11pt,accepted=2026-05-03]{quantumarticle}
\pdfoutput=1
\usepackage{subcaption}
\usepackage{hyperref}
\usepackage[utf8]{inputenc}
\usepackage[english]{babel}
\usepackage[T1]{fontenc}
\usepackage{amsmath}
\usepackage{amsfonts}
\usepackage{amssymb}
\usepackage{amsthm}
\usepackage{mathtools}
\usepackage{float}
\makeatletter
\let\newfloat\newfloat@ltx
\makeatother

\usepackage{algorithm}
\usepackage{algpseudocode}
\usepackage{physics}
\usepackage{relsize}
\usepackage{stmaryrd}
\usepackage{graphicx}
\usepackage{mathrsfs} 
\usepackage{soul}
\usepackage{tikz}
\usetikzlibrary{arrows.meta}

\newtheorem{theorem}{Theorem}

\newtheorem{proposition}{Proposition}
\newtheorem{corollary}{Corollary}

\newtheorem{definition}{Definition}

\newcommand{\setu}{\mathcal{U}}
\newcommand{\setv}{\mathcal{V}}
\newcommand{\setc}{\mathcal{C}}
\newcommand{\sete}{\mathcal{E}}

\newcommand{\setg}{\mathcal{G}}

\begin{document}

\title{Construction and Decoding of Quantum Margulis Codes}

\author{Michele Pacenti}
\affiliation{Department of Electrical and Computer Engineering, University of Arizona, Tucson, AZ, USA, 85721}
\orcid{0009-0002-9648-9041}
\email{mpacenti@arizona.edu}
\thanks{Part of this work is included in the paper ``Quantum Margulis Codes'', that was presented at the 60th Annual Allerton Conference on Communication, Control, and Computing, in Urbana, IL, USA, September 2024. This work is supported by the NSF under grants CCF-2420424, CIF-2106189, CCF-2100013, ECCS/CCSS-2027844, ECCS/CCSS-2052751, and in part by the CoQREATE program under grant ERC-1941583. Bane Vasi\'{c} has disclosed an outside
interest in his startup company Codelucida to The University
of Arizona. Conflicts of interest resulting from this interest are being managed by The University of Arizona in accordance
with its policies.
}

\author{Dimitris Chytas}
\affiliation{Department of Electrical and Computer Engineering, University of Arizona, Tucson, AZ, USA, 85721}
\orcid{0009-0006-7260-9263}

\author{Bane Vasi\'c}
\affiliation{Department of Electrical and Computer Engineering, University of Arizona, Tucson, AZ, USA, 85721}
\orcid{0000-0003-2365-4106}

\maketitle

\begin{abstract}
Quantum low-density parity-check codes are a promising approach to fault-tolerant quantum computation, offering potential advantages in rate and decoding efficiency. In this work, we introduce quantum Margulis codes, a new class of QLDPC codes derived from Margulis' classical LDPC construction via the two-block group algebra framework. We show that quantum Margulis codes, unlike bivariate bicycle codes which require ordered statistics decoding for effective error correction, can be efficiently decoded using a standard min-sum  decoder with linear complexity, when decoded under the code capacity noise model. This is attributed to their Tanner graph structure, which does not exhibit group symmetry, thereby mitigating the well-known problem of error degeneracy in QLDPC decoding. To further enhance performance, we propose an algorithm for constructing 2BGA codes with controlled girth, ensuring a minimum girth of 6 or 8, and use it to generate several quantum Margulis codes of length 240 and 642. We validate our approach through numerical simulations, demonstrating that quantum Margulis codes behave significantly better than BB codes in the error floor region, under min-sum decoding.
\end{abstract}

\section{Introduction}

In the past years, the existence of asymptotically good quantum LDPC codes has been established \cite{panteleev_asymptotically_2022,leverrier_quantum_2022,dinur_good_2023}. However, their practical applicability remains uncertain, particularly in the finite-length regime. Recently, a class of codes known as \textit{bivariate bicycle} (BB) codes \cite{bravyi2024high} was introduced and tested in an experimental fault-tolerant setup. BB codes belong to the broader class of \textit{generalized bicycle} (GB) codes \cite{pryadko2022gb}, which are constructed from pairs of circulant matrices. While GB codes lack the strong asymptotic properties of those in \cite{panteleev_asymptotically_2022,leverrier_quantum_2022,dinur_good_2023}, their suitability for hardware implementation, along with favorable minimum distance and code rate at finite blocklengths, motivates further investigation.

Building on GB codes, a more general framework known as \textit{two-block group algebra} (2BGA) codes was recently proposed \cite{pryadko2024twoblock}. These codes are defined through parity-check matrices derived from the Cayley graphs of certain groups. Notably, GB codes emerge as a special case of this construction. The authors of \cite{pryadko2024twoblock} provided minimum distance bounds and enumerated all 2BGA codes with optimal parameters (up to permutation equivalence) for blocklengths below 200, demonstrating that their minimum distance scales with the square root of the blocklength. The groups considered in these constructions include direct and semidirect products of cyclic and dihedral groups.

We observe that the 2BGA framework naturally extends Margulis' classical LDPC construction \cite{margulis1982explicit} to quantum codes. The classical counterparts of these codes achieve a rate of ${R=1/2}$, with parity-check matrices structured as two square blocks corresponding to the incidence matrices of Cayley graphs of the special linear group $SL(2,\mathbb{Z})$. In this work, we revisit 2BGA codes and introduce their construction through the \textit{left-right Cayley complex}, highlighting their connections to other quantum LDPC families, such as lifted product (LP) codes \cite{panteleev_asymptotically_2022} and quantum Tanner codes \cite{leverrier_quantum_2022}, both of which rely on similar algebraic tools. Through group-theoretic analysis, we demonstrate that, under specific conditions, the group defining a 2BGA code is a subgroup of the automorphism group of its Tanner graph. While this structure has implications for fault-tolerant logical operations \cite{malcolm2025computingefficientlyqldpccodes}, our focus is on its impact on decoding performance. We conjecture that QLDPC codes with weaker symmetries\footnote{By weak symmetry we informally mean that the order of the graph automorphism group is small.} perform better under message-passing decoding. Empirical results support this hypothesis, showing that quantum Margulis codes achieve BB-like error rates without requiring ordered statistics decoding (OSD) when decoding on the depolarizing channel. This allows them to be decoded with linear complexity $\mathcal{O}(n)$, in contrast to BB codes, which typically require $\mathcal{O}(n^3)$. More broadly, this suggests that designing Tanner graphs with a low degree of symmetry could mitigate the performance degradation due to error degeneracy in message-passing decoders. Since decoder performance is also heavily influenced by the Tanner graph's girth, we develop an algorithm to generate 2BGA codes with controlled girth (up to 8). In this paper, we focus on a specific subclass of 2BGA codes, the \textit{quantum Margulis codes} \cite{quantum_margulis_allerton}, and apply our algorithm to construct several instances, including codes of length 240 with girth 6, and length 642 with girth 8. Finally, we evaluate their decoding performance under min-sum (MS) decoding in the code capacity noise model through numerical simulations, and show that they perform better than BB codes when decoded with the same decoder. Finally, we compare the performance of quantum Margulis codes and BB codes under circuit-level noise. All the codes constructed in this paper are available on GitHub \cite{git_repo}. 

The remainder of this paper is organized as follows. In Section \ref{sec:premilinaries}, we provide the necessary preliminaries. Section \ref{sec:2bga} introduces the 2BGA code construction. Section \ref{sec:margulis} presents the construction of quantum Margulis codes, starting with a review of the classical Margulis construction and extending it to the quantum domain. In Section \ref{sec:algorithm}, we describe an algorithm for constructing 2BGA codes with desired girth. Section \ref{sec:decoding} analyzes the decoding of quantum Margulis codes using message-passing algorithms and investigates the impact of graph symmetries on decoding efficiency. Section \ref{sec:performance} provides numerical results on the performance of quantum Margulis codes. Finally, Section \ref{sec:conclusions} concludes the paper with a discussion of key findings and potential directions for future research.

\section{Preliminaries}
\label{sec:premilinaries}

\subsection{CSS codes, Tanner graphs, Cayley graphs}
Let $\mathbb{F}_2^{n}$ be the field of the binary vectors of length $n$; the \textit{Hamming weight} (or simply weight) of an element in $\mathbb{F}_2^{n}$ is the number of its non-zero entries, and we denote it with $|.|$; we will use the same notation to indicate the cardinality of a set, with the intended meaning clear from context. An $[n,k,d]$ linear code $C \subset \mathbb{F}_2^{n}$ is a linear subspace of $\mathbb{F}_2^{n}$ generated by $k$ elements, such that each element in $C$ has Hamming weight at least $d$. A code $C$ can be represented by an $(n-k) \times n$ parity check matrix $\mathbf{H}$ such that $C = \ker \mathbf{H}$. The code rate of $C$ is defined as $R=k/n$. With $C^{\perp} = \mathrm{im}\mathbf{H}^T$ we denote dual of the code $C$ with parameters $[n,k^{\perp}=n-k,d^{\perp}]$. If $\mathbf{H}$ is \textit{sparse}, \textit{i.e.}, its row and column weights are arbitrarily small, the code $C$ is a \textit{low-density parity check} code. A graph $\setg = (\setv,\sete)$ is a collection of vertices $\setv$ and edges $\sete$, such that each edge connects two distinct vertices $v_i,v_j$ and can be represented by the pair $(v_i,v_j)$. A bipartite graph $\setg = (\setu \cup \setv, \sete)$ is a graph where the vertices can be partitioned in two sets, the \textit{left vertices} and the \textit{right vertices}, and the edges only go from one partition to the other. Let $\setu,\setv$ be the sets of left and right vertices, respectively. To a bipartite graph is associated a biadjacency matrix $\mathbf{M} \in \mathbb{F}_2^{|\setu|\times |\setv|}$; each entry of the matrix $m_{i,j}=1$ if there exists an edge between the $i$-th right vertex and the $j$-th left vertex, and $m_{i,j}=0$ otherwise.  The parity check matrix is the biadjacency matrix of the \textit{Tanner~graph} $\mathcal{T}=(\setv\cup \setc,\sete)$, where the nodes in $\setv$ are called \textit{variable nodes} and constitute the set of left vertices, the nodes in $\setc$ are called \textit{check nodes} and constitute the set of right vertices, and there is an edge between $v_j \in \setv$ and $c_i \in \setc$ if $h_{ij}=1$, where $h_{ij}$ is the element in the $i$-th row and $j$-th column of $\mathbf{H}$. The \textit{degree} of a node is the number of incident edges to that node. If all the variable (check) nodes have the same degree we say the code has \textit{regular} variable (check) degree, and we denote it with $d_v$ ($d_c)$. A \textit{cycle} is a closed path in the Tanner graph, and we denote its length by the number of variable and check nodes in the cycle. The \textit{girth} $g$ of a Tanner graph is the length of its shortest cycle. 

Let $(\mathbb{C}^2)^{\otimes n}$ be the $n$-dimensional Hilbert space, and $P_n$ be the $n$-qubit Pauli group; a \textit{stabilizer} group is an Abelian subgroup $S \subset P_n$, and an $\llbracket n,k,d \rrbracket$ stabilizer code  is a $2^k$-dimensional subspace $\mathcal{C}$ of $(\mathbb{C}^2)^{\otimes n}$ that satisfies the condition $s_i\ket{\Psi} = \ket{\Psi},\ \forall\ s_i\in S, \ket{\Psi}\in \mathcal{C}$. An $\llbracket n, k_X-k_Z^{\perp}, d \rrbracket $ CSS code $\mathcal{C}$ is a stabilizer code constructed using two classical  $[n,k_X,d_X]$ and $[n,k_Z,d_Z]$ codes $C_X = \ker \mathbf{H}_X$ and $C_Z = \ker \mathbf{H}_Z$, respectively, such that $C_Z^{\perp} \subset C_X$ and $C_X^{\perp} \subset C_Z$ \cite{calderbank_good_1996}. The minimum distance is $d\geq~\mathrm{min}\{d_X,d_Z\}$, with $d_X$ being the minimum Hamming weight of a codeword in $C_X \setminus C_Z^{\perp}$, and $d_Z$ being the minimum Hamming weight of a codeword in $C_Z \setminus C_X^{\perp}$. Notice that if $\mathbf{H}_X\mathbf{H}_Z^T=\mathbf{0}$ it immediately follows that $C_X^{\perp} \subset C_Z$ and that $C_Z^{\perp} \subset C_X$, thus it is sufficient to satisfy such constraint when designing the two parity check matrices $\mathbf{H}_X$ and $\mathbf{H}_Z$ to obtain a valid CSS code. A quantum CSS code can be seen as a 3-terms \textit{chain complex}. A chain complex 
$$
\cdots \xrightarrow{\partial_{i+1}} \mathcal{C}_i \xrightarrow{\partial_{i}} \mathcal{C}_{i-1} \xrightarrow{\partial_{i-1}} \cdots
$$
is a sequence of abelian groups and morphisms called \textit{boundary maps} such that $\partial_i \circ \partial_{i+1} = 0$ for all $i \in \mathbb{Z}$ \cite{panteleev_asymptotically_2022}. This property implies that $\mathrm{im}\partial_{i+1} \subseteq \mathrm{ker}\partial_i$, thus one can consider the quotient group $H_i(\mathcal{C}) = \mathrm{ker}\partial_i / \mathrm{im}\partial_{i+1}$, called the \textit{i-th homology group} of $\mathcal{C}$. Alternatively, it is possible to define a co-chain complex
$$
\cdots \xleftarrow{\partial^{i+1}} \mathcal{C}^{i+1} \xleftarrow{\partial^{i}} \mathcal{C}^{i} \xleftarrow{\partial^{i-1}} \cdots
$$
to be the dual of a chain complex. Here the morphisms $\partial^i$ are called \textit{co-boundary maps} and $\partial^{i+1} \circ \partial^{i} = 0$, which is equivalent to $\mathrm{im}\partial^{i} \subseteq \mathrm{ker}\partial^{i+1}$, allows us to consider the quotient group $H^i(\mathcal{C}) =  \mathrm{ker}\partial^{i+1} / \mathrm{im}\partial^{i} $. 
A quantum CSS code can be represented by the 3-term chain complex $\mathcal{C}: \mathbb{F}_2^{m_x}\xrightarrow{\partial_2} \mathbb{F}_2^{n} \xrightarrow{\partial_1}  \mathbb{F}_2^{m_z} $, with $\partial_2 \in \mathbb{F}_2^{m_x\times n}$ and  $\partial_1 \in \mathbb{F}_2^{n\times m_z}$ being the first and the second boundary maps, respectively; the space $\mathbb{F}_2^{m_x}$ (resp. the space $\mathbb{F}_2^{m_z}$) corresponds to the space of the $Z$-checks (resp. the $X$-checks), while the space $\mathbb{F}_2^{n} $correspond to the space of the $n$ qubits. Alternatively, we can represent the quantum code by its dual chain $\mathcal{C}^{*}: \mathbb{F}_2^{m_x}\xleftarrow{\partial^1} \mathbb{F}_2^{n} \xleftarrow{\partial^2}  \mathbb{F}_2^{m_z} $. 
It follows that the length of the quantum code is equal to $n$, and its dimension $k$ is the dimension of the first homology group $H_1(\mathcal{C})$ (or of its first cohomolgy group $H^1(\mathcal{C^*})$).
We naturally identify $\mathbf{H}_X \triangleq \partial^1$ and  $\mathbf{H}_Z \triangleq \partial_1$. The logical codewords, also known as \textit{logical operators} of the CSS code are the elements of $H_1(\mathcal{C})$ or $H^1(\mathcal{C^*})$ (the two are isomorphic).


Given a group $G$ and a set of generators $S$, it is possible to construct the Cayley graph $\mathrm{Cay}(G,S)$ of $G$ with respect of $S$, such that $\mathrm{Cay}(G,S) = (G,\mathcal E_S)$ is a graph where there is a vertex for every element $g \in G$, and there is an edge for every pair $(g,gs)$, with $s\in S$, if $S$ acts on the right. Alternatively, if $S$ acts on the left, the edges have the form $(g,sg)$.
For this paper we will only consider bipartite Cayley graphs, obtained as double covers of Cayley graphs. In a bipartite Cayley graph $\mathrm{Cay}(G\times \{0,1\},S)$ the sets of left and right vertices are two copies of $G$, and \emph{every} edge is of the form
$$
((g,0),(gs,1)) \qquad \text{for all } g\in G,\; s\in S,
$$
with the two sets of vertices corresponding to $G\times\{0,1\}$.
If $S=S^{-1}$, then the graph is undirected, meaning that each edge $((g,0),(gs,1))$ is paired with its reverse $((gs,0),(g,1))$. 
Thus the bipartite Cayley graph is undirected if and only if $S$ is inverse-closed.

\subsection{Error model and decoding}
In this paper, we consider two error models for our numerical simulations. The first is the code capacity model, where each physical qubit has a probability of undergoing through an $X$, $Z$ or $Y$ error with probability $\epsilon/3$, and with probability $1-\epsilon$ has no error. The second error model is the circuit-level noise model, where we simulate the syndrome extraction circuit of the code, such that each operation in the circuit has a probability $p$ of being faulty \cite{higgott_improved_2023}. A faulty qubit initialization prepares a single-qubit state orthogonal to the correct one with probability $p$. A faulty CNOT is an ideal CNOT followed by one of the 15 non-identity Pauli errors on the control and the target qubits, picked uniformly at random with probability $p/15$. A faulty measurement is an ideal measurement followed by a classical bit-flip error applied to the measurement outcome with probability $p$. An idling qubit suffers from depolarizing noise, with depolarizing probability $\epsilon = p$. Because measurements are noisy, we utilize the well-known repeated measurements strategy, and utilize detectors associated with measurements: let $m_1,...,m_R$ the value of a particular check at each measurement round, with $R$ being the total number of rounds. The resulting detectors associated with the same check will be $m_1, m_1\oplus m_2, m_2\oplus m_3,...,m_{R-1}\oplus m_R$ \cite{gong2024lowlatencyiterativedecodingqldpc}. For our circuit, we utilize a measurement schedule perfectly equivalent to the one in \cite{bravyi2024high}. The decoder runs on the so-called \textit{circuit-level} Tanner graph \cite{higgott_improved_2023}, where each variable node is an independent error mechanism, and each check node is a detector.

A Pauli error $E \in P_n$ can be decomposed into $\mathbf{e}_X,\mathbf{e}_Z \in \mathbb{F}_2^n$, such that $e_{X_i}=1$ if $E_i=\{X,Y\}$ and $e_{X_i}=0$ otherwise, and $e_{Z_i}=1$ if $E_i=\{Z,Y\}$ and $e_{Z_i}=0$ otherwise. Given $\mathbf{e}_X$ and $\mathbf{e}_Z$, the binary syndrome $\mathbf{s} = [\mathbf{s}_X, \mathbf{s}_Z]$ can be obtained in CSS codes such that $\mathbf{s}_X = \mathbf{e}_X\mathbf{H}_Z^T$ and $\mathbf{s}_Z = \mathbf{e}_Z\mathbf{H}_X^T$, and correction of $X$ and $Z$ errors can be carried out separately. 
If $E$ belongs to the stabilizer group $S$, by definition it will commute with all the stabilizers. In CSS codes this can be seen by observing that any row $\mathbf{h}_j \in \mathbf{H}_X$ is orthogonal to all the rows of $\mathbf{H}_Z$ (and vice-versa). This means that every error $\mathbf{e}_X + \mathbf{h}_j$, with $\mathbf{h}_j \in \mathbf{H}_X$ and for all $j$, will correspond to the same syndrome, but since stabilizers do not affect the code space, any error estimate $\hat{\mathbf{e}} = \mathbf{e}_X + \mathbf{h}_i$, with $i$ equal or different from $j$, is a valid error estimate (the same holds for $\mathbf{e}_Z$ and $\mathbf{H}_Z$). Given an error $\mathbf{e}_X$, we identify the set of \textit{degenerate errors} all the errors of type $\mathbf{e}_X + \mathbf{h}_j$. Clearly, all the degenerate errors are associated to the same syndrome, and they form a coset. Although, in principle, error degeneracy may facilitate the decoding process, as for a given error there are several valid estimates, in practice it often constitutes an obstacle, especially for iterative decoders that may exhibit oscillatory behavior, failing to converge to a valid error estimate \cite{chytas2025enhancedminsumdecodingquantum,chytas_enhanced_2024,chytas2024collectivebitflippingbaseddecoding}.

A message-passing decoder is an algorithm operating on the Tanner graph of the code in an iterative fashion. The input of the decoder is a measured syndrome, while the output is an estimate of the error that generated the syndrome. In each iteration, messages between variable nodes and check nodes are exchanged. We denote the variable-to-check message from the variable node $j$ to the check node $i$ at the $\ell$-th iteration with $\nu_{j,i}^{\ell}$, and the check-to-variable message from check node $i$ to variable node $j$ at the $\ell$-th iteration with $\mu_{i,j}^{\ell}$. In this paper, we consider two-message passing decoders: the belief propagation (BP) \cite{sum-product} and the min-sum (MS) \cite{nms_decoder}. Let $\lambda_j$ be the \textit{a priori} log-likelihood ratio for the $j$-th variable node, obtained as $\lambda_j = \log (\frac{1-\alpha_j}{\alpha_j})$, where $\alpha_j$ is the prior probability of error for that qubit. In both the decoders the variable-to-check message is computed as:
\begin{equation}
    \nu_{j,i}^{\ell} = \lambda_j + \sum_{i' \in \mathcal{N}(j)\setminus i} \mu_{i',j}^{\ell},
\end{equation}
where $\mathcal N(j)$ indicates the set of indices of the neighbors of the node labeled $j$.
In the BP decoder, the check-to-variable message is computed as:
{\smaller
\begin{equation}
    \mu_{i,j}^{\ell} = 2(1-2s_i)\cdot \mathrm{tanh}^{-1}\left\{\prod_{j' \in \mathcal{N}(i) \setminus j} \mathrm{tanh}\left\{\frac{\nu_{j',i}^{\ell-1}}{2}\right\}\right \}.
\end{equation}
}

In the MS decoder, the check-to-variable message is an approximated version of the one of BP, which is faster to compute in practice:
\begin{equation}
    \mu_{i,j}^{\ell} = (1-2s_i)\cdot \prod_{j' \in \mathcal{N}(i) \setminus j} \mathrm{sgn}(\nu_{j',i}^{\ell-1})\cdot  \min |\nu_{j',i}^{\ell-1}|,
\end{equation}
where the sign function is defined as
\begin{align}
    \mathrm{sgn}(x) = & \begin{cases}
        -1\ \mathrm{if}\ x<0\\
        +1\ \mathrm{otherwise}.
    \end{cases}
\end{align}
In this work we also consider the normalized version of the MS decoder, denoted with nMS, where check-to-variable messages are multiplied by a scalar $\beta \in [0,1]$. The nMS decoder is commonly used in classical error correction where it provides a good approximation of the BP decoder if the normalization parameter is carefully selected \cite{nms_decoder}. The error estimate $\hat{\mathbf{e}}$ is given by the hard decision on the soft information of the variable nodes, such that
\begin{equation}
    \hat{e}_j = \mathrm{sgn}\left(\lambda_j + \sum_{i \in \mathcal{N}(j)} \mu_{i,j}^{\ell} \right).
\end{equation}
The decoding procedure continues until the maximum number of iterations has been reached, or until the error estimate $\hat{\mathbf{e}}$ matches the input syndrome.

\section{Two-block group-algebra code construction}
\label{sec:2bga}

In this Section, we illustrate the construction proposed in \cite{pryadko2024twoblock} throughout the framework of the left-right Cayley complex, a combinatorial tool utilized in \cite{panteleev_asymptotically_2022,leverrier_quantum_2022,dinur_good_2023}.
A left-right Cayley complex is constructed in the following manner. Let $G$ be a group and $A,B \subset G$ two sets of generators, $A$ acting on the left and $B$ acting on the right. We take 4 copies of $G$, namely 
$G\times \{0,1\}^2$
(with elements written as $(g,ij)$ with $i,j\in \{0,1\}$). 
For each element $(g,00) \in G\times \{0,0\}$, and for every pair $(a,b) \in A\times B$, we construct the square 
$$
\{(g,00), (bg,01), (ga,10), (bga,11)\}.
$$  
Each set $G\times \{i,j\}$ constitutes a set of vertices, while elements of $A$ and $B$ are the edges of the complex. The group structure assures that each vertex has $|A|+|B|$ incident edges, and $|A|\times|B|$ incident squares.

\begin{figure}
\centering
\begin{minipage}[t]{0.48\textwidth}
\centering
\input{figures/square.tikz}
\subcaption{}
\label{sufig1}
\end{minipage}%
\hfill%
\begin{minipage}[t]{0.48\textwidth}
\centering
\input{figures/square2.tikz}
\subcaption{}
\label{sufig2}
\end{minipage}
\caption{(a) A square associated with the chain complex $X$. (b) A square associated with the co-chain complex $X^*$. }
\label{fig:square}
\end{figure}
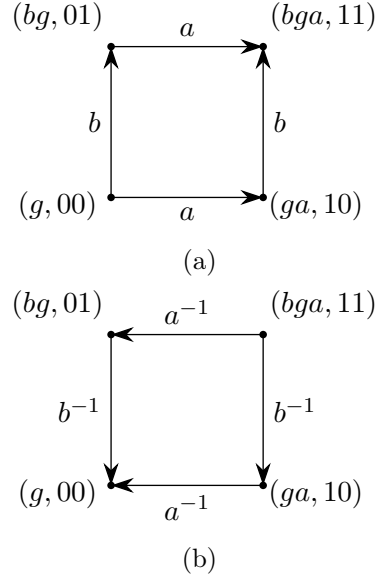

A single instance of square is represented in Fig.~\ref{fig:square}; notice that the subgraphs 
$((G,00)\cup (G,10),A)$
and 
$((G,01)\cup (G,11),A)$
are two copies of $\mathrm{Cay}(G\times \{0,1\},A)$, and similarly, the subgraphs 
$((G,00)\cup (G,01),B)$
and 
$((G,10)\cup (G,11),B)$
are two copies of $\mathrm{Cay}(G\times \{0,1\},B)$. 
To construct our complex, we associate $X$ stabilizers with the vertices in $G\times \{0,0\}$, $Z$ stabilizers with the vertices in $G\times \{1,1\}$, and qubits with vertices in 
$G\times \{1,0\} \cup G\times \{0,1\}$
\cite{quantum_margulis_allerton}. 
We then construct the three terms chain complex over the group algebra $\mathbb{F}_2G$:
\begin{equation}
        X : \mathbb{F}_2^{G_{00}} \xrightarrow{\partial_2} \mathbb{F}_2^{G_{01}} \oplus \mathbb{F}_2^{G_{10}} \xrightarrow{\partial_1}  \mathbb{F}_2^{G_{11}},
\end{equation}
where with $\mathbb{F}_2^G$ we denote the space of all the formal linear combinations of elements of $G$ with binary coefficients.
Here, the boundary maps are respectively $\partial_2=[\mathbf{A},\mathbf{B}]$, where $\mathbf{A},\mathbf{B} \in \mathbb{F}_2^{|G|\times |G|}$ are the biadjacency matrices of the double covers of the Cayley graphs of $G$ in respect of $A$ and $B$, respectively, and $\partial_1 = [\mathbf{B},\mathbf{A}]^T$.
By defining $A$ and $B$ to operate on the right and on the left, respectively, we obtain that $\mathbf{A}\mathbf{B} = \mathbf{B} \mathbf{A}$, thus the chain complex is properly defined, as $\partial_2 \circ \partial_1 = [\mathbf{A},\mathbf{B}]\cdot [\mathbf{B},\mathbf{A}]^T=\mathbf{A}\mathbf{B}+\mathbf{B}\mathbf{A}=\mathbf{0}$. Similarly, one can define the co-chain complex
\begin{equation}
            X^* : \mathbb{F}_2^{G_{00}} \xleftarrow{\partial^1} \mathbb{F}_2^{G_{01}} \oplus \mathbb{F}_2^{G_{10}} \xleftarrow{\partial^2}  \mathbb{F}_2^{G_{11}}, 
\end{equation}
such that $\partial^1 = [\mathbf{A}^T,\mathbf{B}^T]^T$ and $\partial^2 = [\mathbf{B}^T,\mathbf{A}^T]$. Hence, the parity check matrices of the code will have the form
\begin{equation}
\label{eq:matrices}
\begin{cases}
    \mathbf{H}_X = & \partial_2 = [\mathbf{A}\ \mathbf{B}] \\ 
    \mathbf{H}_Z = & \partial^2 = [\mathbf{B}^T\ \mathbf{A}^T] 
\end{cases}
\end{equation}

Note that $\mathbf{A}^T,\mathbf{B}^T$ correspond to the biadjacency matrices of the double covers of the Cayley graphs of $G$ in respect of $A^{-1}$ and $B^{-1}$, respectively; indeed, we can represent the co-chain complex $X^*$ by reversing the arrows, as illustrated in Fig. \ref{sufig2}, and substituting $a$ and $b$ with $a^{-1}$ and $b^{-1}$. Notice that the corresponding Cayley graphs are directed, thus biadjacency matrices may not be symmetric.

\subsection{Automorphism group of 2BGA codes}
Because of the rich structure of 2BGA codes, it is possible to make statements on the Tanner graph automorphism of such codes. Let us first introduce the notion of $G$-invariant graph.

\begin{definition}
\label{def:action}
 A bipartite graph $\setg = (\setu\cup \setv,\sete)$ is $G$-invariant if there exist $G$-actions on $\setu$ and $\setv$ such that if $(u,v) \in \sete$, then $(ug,vg) \in \sete$. This is equivalent to say that the graph has $G$-symmetry. If $\forall\ u_1,u_2$ there exists $g\in G$ such that $u_1g=u_2$, then the action of $G$ is transitive. Moreover, if $u_1g=u_1 \iff g=e$, where $e$ is the identity, then the action of $G$ is free.
\end{definition}

Obviously, the Cayley graph of a group $G$ is $G$-invariant. Because of the structure of $\mathbf{H}_X,\mathbf{H}_Z$, there is a natural one-to-one mapping between the set of check nodes of the Tanner graphs $\mathcal{T}_X,\mathcal{T}_Z$, and the elements of $G$.
Using this fact, we prove that, under certain conditions, the group $G$ utilized to construct the 2BGA code is also a subgroup of the automorphism group of the Tanner graphs of the code.

\begin{theorem}
\label{th:invariance}
    Let $\mathcal{T} = (\setv \cup \setc,\sete)$ be the $X$ (or $Z$) Tanner graph of a 2BGA code obtained from a group $G$ and two sets of generators $A,B$. Then, if $A$ ($B$) is a normal subgroup of $G$, $G \subseteq \mathrm{Aut}(\mathcal{T})$, assuming it to act on the right (left).
    \begin{proof}
        Consider a check node labeled as $g \in G$ and the two edges $(g,ga)$ and $(g,bg)$. If $\mathcal{T}$ is $G$-invariant, assuming $G$ to act on the right, for any $h\in G$ we have that the edges $(gh,gah)$ and $(gh,bgh)$ must exist. Clearly, since the check $gh$ must exist, the edge $(gh,bgh)$ is obtained by multiplying $gh$ by $b$ on the left, therefore such edge exists. On the other hand, if $(gh,gah)$ exists it means there exists $x\in A$ such that $ghx = gah$, meaning that $x=h^{-1}ah$. If $A$ is a normal subgroup of $G$, it is invariant under conjugation, \textit{i.e.}, for any $h \in G$ we have $h^{-1}Ah=A$, thus $x\in A$ and the edge exists. Assuming the action of $G$ to be from the right, the proof is analogous and it requires $B$ to be normal.
    \end{proof}
\end{theorem}

\begin{corollary}
    Let $\mathcal{T} = (\setv \cup \setc,\sete)$ be the $X$ (or $Z$) Tanner graph of a 2BGA code obtained from a group $G$ and two sets of generators $A,B$. If $G$ is Abelian, $G \subseteq \mathrm{Aut}(\mathcal{T})$.
    \begin{proof}
    If $G$ is Abelian, all its subgroups are normal, therefore the hypothesis for Theorem \ref{th:invariance} is always satisfied.
    \end{proof}
    \label{th:abelian}
\end{corollary}

It is evident that all BB codes exhibit cyclic symmetry, as the cyclic group is Abelian. However, Theorem \ref{th:invariance} establishes that for a non-Abelian group \( G \), symmetry is preserved only if one of the generating sets, \( A \) or \( B \), forms a normal subgroup of \( G \). Since this condition is rarely met in practice, it follows that for non-Abelian \( G \), the associated 2BGA code generally lacks \( G \)-symmetry. As we will demonstrate, this absence of symmetry enables more efficient decoding through message-passing algorithms. We will later utilize the following well-known property of graphs with $G$-symmetry.

\begin{proposition}
\label{proposition:symmetry}
    Let $\mathcal{T} = (\setv \cup \setc,\sete)$ be the $X$ (or $Z$) Tanner graph of a 2BGA code obtained from a group $G$ and two sets of generators $A,B$. Assume $G$ to be Abelian, so that $\mathcal T$ is $G$-invariant and $G \subseteq \mathrm{Aut}(\mathcal{T})$. Then, for any subgraph $t \in \mathcal{T}$ there exist at least $|G|$ isomorphic copies of $t$.
    \begin{proof}
        Each edge in $t$ has the form $(g_1,g_2)$, with $g_1,g_2 \in G$. Since $\mathcal{T}$ is $G$-invariant, we have that for any $h \in G$, the edge $(g_1h,g_2h)$ also exists. Thus, it naturally follows that for any subgraph $t$, one can obtain an isomorphic subgraph $t'$ by multiplying each vertex by~$h \in G$. 
    \end{proof}
\end{proposition}
If we apply Proposition \ref{proposition:symmetry} to a check node and its neighborhood up to a certain depth, we obtain that each check node in $\mathcal{T}$ has an isomorphic neighborhood to all the others. We will utilize this fact later in Section \ref{sec:decoding}.

\section{Quantum Margulis Codes}
\label{sec:margulis}
\subsection{Classical Margulis construction}
Margulis codes \cite{margulis1982explicit} are a well-known class of classical LDPC codes constructed from Cayley graphs of certain groups. 
Let $G$ be $SL(2,p)$ the \textit{Special Linear Group} whose elements consist of $2\times 2$ matrices of determinant 1 over $\mathbb{Z}_p$, being $p$ prime. Let $S$ be a set of generators of $SL(2,\mathbb{Z}_p)$ chosen according to the construction we report below, and let $S^{-1}$ be the inverse of $S$. Let $\setg_0 = \mathrm{Cay}(G,S)$ be the bipartite Cayley graph of $G$ with respect of $S$ and $\setg_1 = \mathrm{Cay}(G,S^{-1})$ be the bipartite Cayley graph of $G$ with respect of $S^{-1}$; consider $\mathbf{M}_0,\mathbf{M}_1$ the biadjacency matrices of $\setg_0,\setg_1$, respectively. The parity check matrix of the Margulis code has the form
\begin{equation}
    \mathbf{H} = [\mathbf{M}_0\  \mathbf{M}_1].
\end{equation}
The resulting code has blocklength $n = 2(p^2-1)p$, rate $R=1/2$, variable degree $d_v=|S|$ and check degree $d_c=2|S|$.
A similar construction is based on Ramanujan graphs by Lubotzky \textit{et al.} \cite{rosenthal2000,MACKAY200397}, however we will only consider Margulis construction for this paper.

Margulis shows that if the set of generators are chosen such that there is no multiplicative relation between them, the girth of the graph grows as $\log p$; moreover, he gives an explicit construction of the generating set for any degree satisfying this property. Let $\eta$ be a sufficiently large integer, and let us select $r+1$ distinct pairs $(m_i,q_i)$, with $1 \leq i \leq r+1$, such that $\gcd (m_i,q_i) = 1$ and $0 \leq m_i \leq \eta/2$, $0 \leq q_i \leq \eta/2$. For each pair, there exist a matrix
\begin{equation}
    C_i = \begin{pmatrix}
        m_i & a_i \\
        q_i & b_i
    \end{pmatrix} \in SL(2,\mathbb{Z})
\end{equation}
such that $|a_i|,|b_i| < \eta/2$. Each generator $g_i$ is then given by 
\begin{equation}
    g_i = C_i \begin{pmatrix}
        1 & \eta \\
        0 & 1
    \end{pmatrix}C_i^{-1}, \forall\ i=1,..,r+1.
\end{equation}
The generating set is defined as $S = \{g_1,...,g_r\}$, with $S^{-1}=\{g_1^{-1},...,g_r^{-1}\}$. Because each $g_i \in SL(2,\eta\mathbb{Z})$, Margulis shows that there exist no nontrivial multiplicative relation between the generators, and that the girth of the code grows as $\mathcal{O}(\log n / \log r)$.

\subsection{Quantum Margulis construction}
The construction from Margulis can be extended to fit the design proposed in Section \ref{sec:2bga}. Let $G=SL(2,\mathbb{Z}_p)$, with $p$ prime. In principle, one can use Margulis' method to obtain two sets $A,B$ of $r$ generators each and their inverses $A^{-1},B^{-1}$. Let $A$ act on the left and $B$ act on the right; we construct the left-right Cayley complex as in Section \ref{sec:2bga}, and the associate the quantum code. The parity check matrices $\mathbf{H}_X,\mathbf{H}_Z$ have the form illustrated in (\ref{eq:matrices}).

The quantum code has length \mbox{$n=2(p^2-1)p$}, regular variable degree $d_v=r$, and regular check degree $d_c=2r$. Because both $\mathbf{H}_X$ and $\mathbf{H}_Z$ have $|G|$ rows and $2|G|$ columns, the rate of the quantum code $R \rightarrow 0$ for $n\rightarrow \infty$; however, because they always have redundant rows, we are able to construct finite length codes with non-trivial dimension\footnote{Note that this generally depends on the specific choice of the group and of its generating set. In general, the code has trivial dimension. GB codes also have vanishing rate.}.
We notice that there are few differences between classical and quantum Margulis codes: first, the girth of the classical Margulis codes is shown to increase as $\mathcal{O}(\log n / \log r)$, while for the quantum Margulis code it is bounded by $g\leq8$. This is a well-known fact about HP and LP codes, as well as for 2BGA codes \cite{raveendran_trapping_2021}, and it is due to the presence of stabilizers combined with $d_v\geq 3$. Although there are 8-cycles that cannot be removed from the Tanner graph, one can still eliminate 4-cycles and 6-cycles.
Notice that Margulis' argument on the girth still holds inside each separate block $\mathbf{A},\mathbf{B}$, thus one can use his construction to obtain codes with less short cycles while increasing the blocklength. However, in this paper we construct our codes using a randomized algorithm, that can be adopted to construct any 2BGA code without short cycles.

\section{Constructing 2BGA codes with high girth}
\label{sec:algorithm}

The girth of the Tanner graph plays a crucial role in the performance of iterative message-passing decoders. It is well known that decoding effectiveness is closely tied to both the length and number of cycles in the Tanner graph. In classical LDPC code design, significant effort is devoted to eliminating short cycles to improve decoding performance.  In this paper, we propose an algorithm that, given a group \( G \), selects the generators of the code to ensure the resulting Tanner graph avoids four-cycles and six-cycles. The algorithm achieves this by iteratively expanding a root node, which, by convention, corresponds to the identity element of \( G \). 
The procedure is formalized in Algorithm \ref{alg:obtain_generators} and Algorithm \ref{alg:generate_tree}. In Algorithm \ref{alg:obtain_generators}, a random generator set is initially selected, followed by a call to Algorithm \ref{alg:generate_tree} to construct the first layer of the Tanner graph. Specifically, each element \( l \in  \mathcal{L}_{\ell} \) (the $\ell$-th layer) is multiplied by each element of \( S \), with the exception of the one that generated it in the previous layer ($a_l)$. If \( \ell \) is even, \( \mathcal{L}_{\ell+1} \) consists of check nodes, while for odd \( \ell \), two distinct sets, \( \mathcal{L}_R \) and \( \mathcal{L}_L \), correspond to variable nodes from the two disjoint blocks $\mathbf{A}$ and $\mathbf{B}$.

\begin{algorithm}
    \caption{$\mathtt{get\_generators}$}
    \label{alg:obtain_generators}
    \begin{algorithmic}[1]
        \Require A group $G$, the size of the generating set $w$ and a target girth $\gamma$
        \Ensure A set of generators $S$ that generate a code with girth $\geq \gamma$
         \input{Algorithms/get_marg_gen}
    \end{algorithmic}
\end{algorithm}

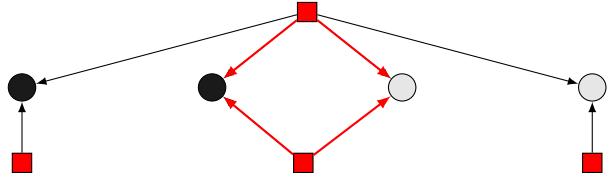
\begin{figure}
\centering
\begin{minipage}[t]{0.48\textwidth}
\centering
\input{tikz/tree1}
\subcaption{}
\label{fig:four-cycle}
\end{minipage}%
\hfill%
\begin{minipage}[t]{0.48\textwidth}
\centering
\input{tikz/tree2}
\subcaption{}
\label{fig:six-cycle}
\end{minipage}
\caption{(a) A 4-cycle in the expanded neighborhood of a check with $d_c=4$. Red square nodes indicate checks, while circles indicate variable nodes. Black and gray variable nodes indicate qubits belonging to the two blocks $\mathbf{A}$ and $\mathbf{B}$. (b) A 6-cycle in the expanded neighborhood of a check with $d_c=4$.}
\label{fig:expanded-neighborhood-cycles}
\end{figure}

In this case, elements in \( \mathcal{L}_{\ell} \) are multiplied by the inverses of \( A \) and \( B \) since the graph is (in general) directed from check nodes to variable nodes, as mentioned in Section \ref{sec:2bga}. To find a cycle is sufficient to check that all the elements in $\mathcal{L}_{\ell +1}$ are distinct. If not, it means that two or more edges are incident to the same node, and a flag $f$ is set so that Algorithm \ref{alg:obtain_generators} changes one of the generators involved in the cycle (stored in the variable $\setc$). If no cycles up to depth $\gamma/2$ are found, and $G$ is Abelian, the 2BGA code constructed by the generators $S$ is guaranteed to have girth of at least $\gamma$, thanks to Proposition \ref{proposition:symmetry}. Otherwise, the neighborhood of each check node $g \in G$ is constructed using Algorithm \ref{alg:obtain_generators}, and if a cycle is found, the Algorithm starts again from the first node by choosing a different set of generators.
Figures \ref{fig:four-cycle} and \ref{fig:six-cycle} illustrate examples of Tanner graphs generated by Algorithm \ref{alg:obtain_generators}, highlighting cases where cycles appear.  We use black and gray to highlight variable nodes belonging to the two separate blocks $\mathbf{A}$ and $\mathbf{B}$ of a code with $d_v=2$ and $d_c=4$, and red for check nodes. The edges of the graph are directed, and we highlight in red the edges that form a cycle.

\begin{algorithm}
    \caption{$\mathtt{generate\_tree}$}
    \label{alg:generate_tree}
    \textbf{Input:} $S$, $\mathcal{L}_{\ell}$, $\gamma$, $\ell$\\
    \textbf{Output:} Generators $S$, flag $f$, cycle detected $\setc$, depth $\ell$.
    \input{Algorithms/generate_tree}
\end{algorithm}

\section{Decoding of quantum Margulis code}
\label{sec:decoding}
\begin{figure}
    \centering
    \includegraphics[scale=0.7, trim=2.5cm 1.3cm 1.5cm 0.9cm, clip=true]{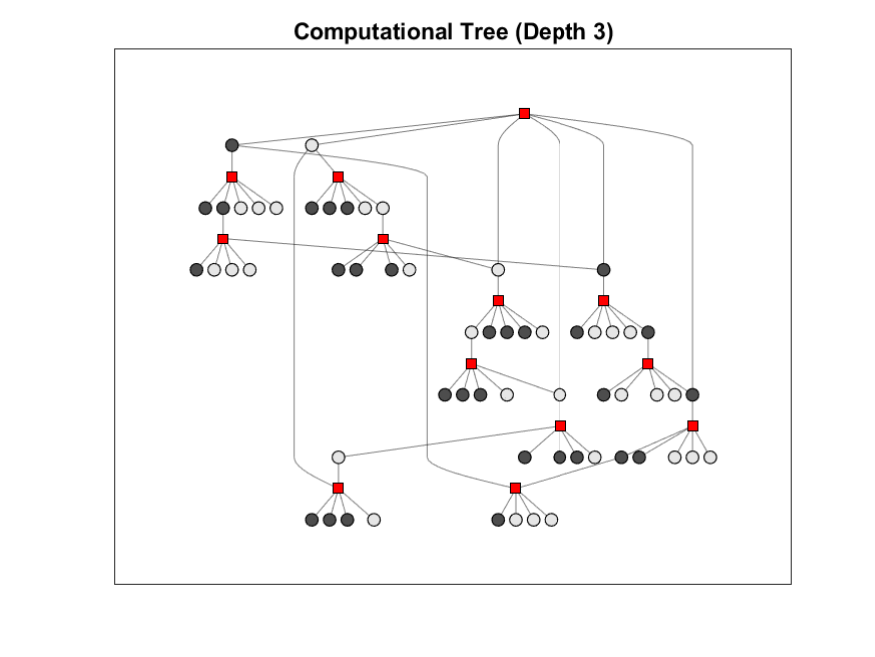}
    \caption{Depth-3 neighborhood of a check node in the symmetric stabilizer in the $\llbracket 72,12,6 \rrbracket$ BB code. Due to the graph's symmetry, every check node in the graph has this neighborhood.}
    \label{fig:bbcode_neigh}
\end{figure}

In this Section we analyze the performance of quantum Margulis codes under iterative decoding. In particular, we show that a simple nMS decoder with flooding schedule is able to approach the BPOSD performance. 
First, let us recall what are the issues in message-passing decoders for QLDPC codes, and why OSD post-processing is required. It is well-known that message passing decoders such as BP or MS fail while correcting errors on QLDPC codes due to the code's degeneracy. In particular, it has been shown that error degeneracy can be associated to particular structures in the Tanner graph called \textit{quantum trapping sets} (qTS), also known as \textit{symmetric stabilizers} \cite{raveendran_trapping_2021}.
\begin{definition}
    A symmetric stabilizer is a stabilizer with the set of variable/qubit nodes, whose induced sub-graph has no odd-degree check nodes, and that can be partitioned into an even number of disjoint subsets, so that: (a) sub-graphs induced by these subsets of variable nodes are isomorphic, and (b) each subset has the same set of odd degree check node neighbors in its induced sub-graph.
\end{definition}
\begin{figure}
    \centering
    \input{tikz/symmstab.tikz}
    \caption{A symmetric stabilizer for a $d_v=3$ 2BGA code.}
\end{figure}
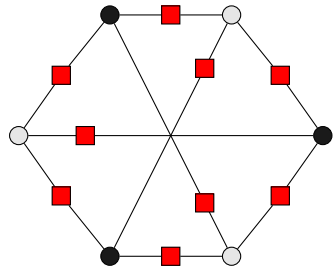
The behavior of message-passing decoders when errors fall into a symmetric stabilizer has been widely studied \cite{chytas_enhanced_2024,chytas2024collectivebitflippingbaseddecoding,chytas2025enhancedminsumdecodingquantum}: in particular, it has been observed that message-passing decoders typically fail when half-weight error patterns are applied to symmetric stabilizers as they are pushed to converge to both degenerate error-patterns inside the stabilizer.
In particular, in \cite{chytas2025enhancedminsumdecodingquantum} it has been proved that under the isolation assumption, the variable-to-check messages sent by MS inside the symmetric stabilizer can be reduced to a linear dynamical system, and that they oscillate for any half-weight error pattern applied on the stabilizer. In such scenario, the isolation assumption is reasonable, because the messages incoming from the external graph are perfectly symmetric, thus they do not contribute to the decoder's convergence.
As an example, in Fig.~\ref{fig:bbcode_neigh} we illustrate the depth-3 neighborhood of a check node in the $\llbracket 72,12,6 \rrbracket$ BB code; it is easy to see that in such code, the neighborhood structure of every check node is isomorphic to the neighborhood structure of all the others, as shown in Proposition \ref{proposition:symmetry}. This holds for all the 2BGA codes obtained from an Abelian group.

\begin{figure*}
    \centering
    \begin{minipage}{0.33\textwidth}
        \centering
            \includegraphics[width=1\linewidth,trim=3.2cm 1.3cm 1.5cm 0.9cm, clip]{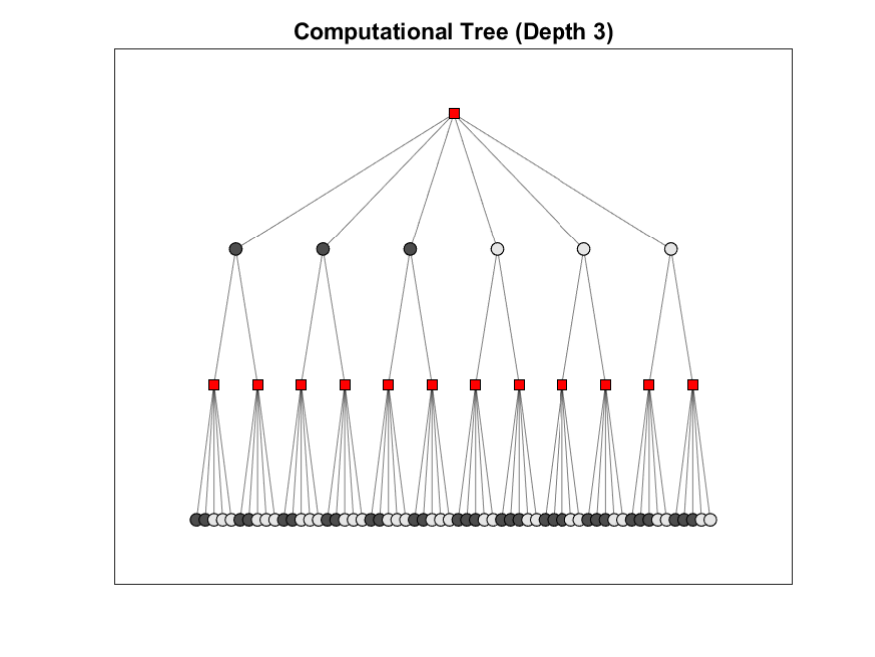}
    \subcaption{}
    \end{minipage}%
    \begin{minipage}{0.33\textwidth}
        \centering
        \includegraphics[width=1\linewidth,trim=3.2cm 1.3cm 1.5cm 0.9cm, clip]{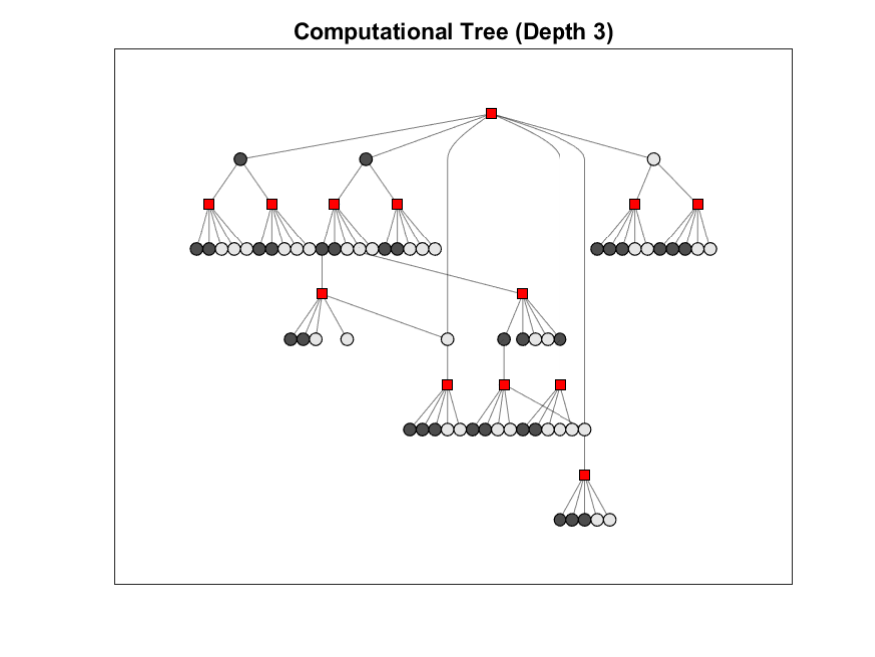}
        \subcaption{}
    \end{minipage}%
        \begin{minipage}{0.33\textwidth}
        \centering
            \includegraphics[width=1\linewidth,trim=3.2cm 1.3cm 1.5cm 0.9cm, clip]{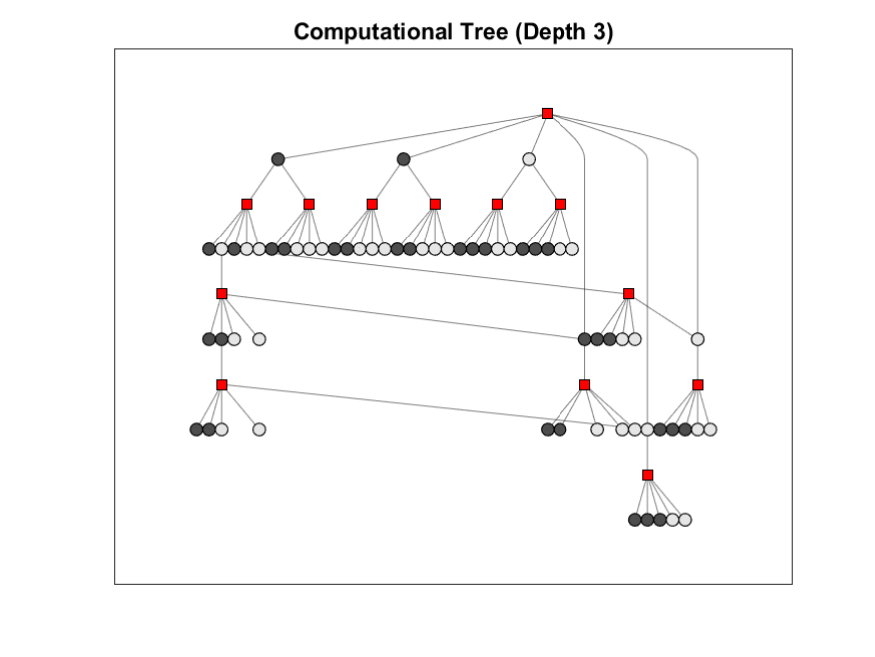}
        \subcaption{}
    \end{minipage}
    \caption{Depth-3 neighborhood of three different check nodes of a symmetric stabilizer of the $\llbracket 240,2 \rrbracket$ quantum Margulis code. While it is evident that in (a) the structure is a tree, one can observe that the topologies in (b) and (c) are also different. For instance, in (b) we can distinguish a 6-cycle and an 8-cycle, while in (c) there are two 6-cycles and a 10-cycle.}
    \label{fig:marg_neigh}
\end{figure*}

The solution proposed in \cite{chytas2025enhancedminsumdecodingquantum} to solve the symmetry problem is to adopt a different update rule for the decoder, only for the black (or the gray) variable nodes, which would introduce asymmetry in the messages of the graph. However, we notice that the nMS decoder does not get stuck in the symmetric stabilizers of quantum Margulis codes. We conjecture that this effect is due to the fact that the group utilized to construct the codes, namely $SL(2,\mathbb{Z}_p)$, is non-Abelian, meaning that the Tanner graph of the code does not show the symmetry described by Theorem \ref{th:invariance}. This means that the neighborhood structure of each check node is, in principle, different, as we show in the three examples of Fig \ref{fig:marg_neigh}. This property ensures that, after a sufficient amount of decoding iterations, the messages introduced in the symmetric stabilizer from its check nodes are asymmetric, thus leading the decoder to converge to one of the possible degenerate errors.

\begin{figure}
    \centering
    \includegraphics[width=.5\textwidth]{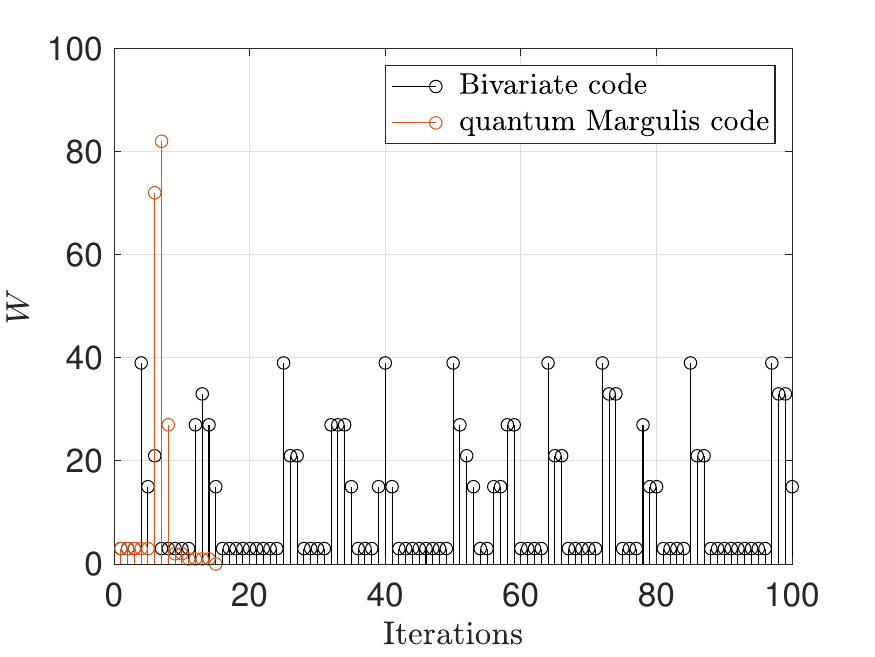}
    \caption{Evolution of the weight of the estimated error in the MS decoder.}
    \label{fig:weight-iter}
\end{figure}

We provide numerical results to support our claim. In particular, we insert errors into symmetric stabilizers of a quantum Margulis code and a BB code, and observe the behavior of the decoder. In Fig. \ref{fig:weight-iter} we illustrate the quantity $W_k = |\hat{\mathbf{e}}_k| + |\mathbf{e}| + |\mathbf{s}|$, where $\mathbf{s}$ is the support of a stabilizer, $\mathbf{e} \subset \mathbf{s}$ is the support of the error, and $\hat{\mathbf{e}}_k$ is the support of the error estimate at the $k$-th iteration. We choose $\mathbf{e}$ to be of weight half of the stabilizer weight. On the $x$ axis we plot the number of iterations. The figure highlights how the decoder is able to converge after 15 iterations on the quantum Margulis code (where $W$ goes to 0), where for the BB code it gets stuck in a local minima with $W=3$.

\begin{figure}
    \centering
    \includegraphics[width=.5\textwidth]{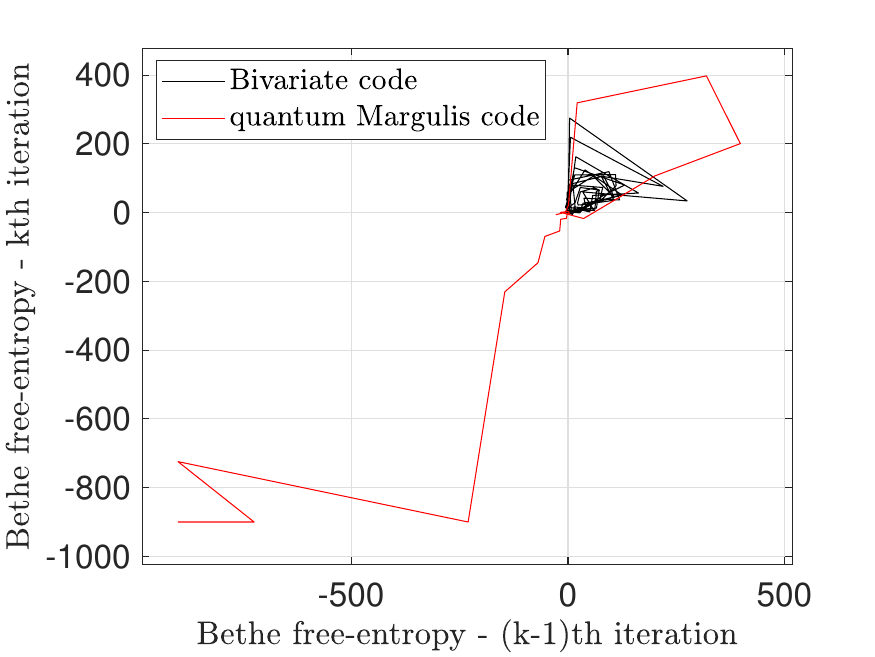}
    \caption{Trajectory of the Bethe free-entropy function (\ref{eq:entropy}) of the nMS decoder while decoding a symmetric stabilizer on a BB code and a quantum Margulis code.}
    \label{fig:bethe}
\end{figure}

\begin{figure}
    \centering
    \includegraphics[width=0.5\textwidth]{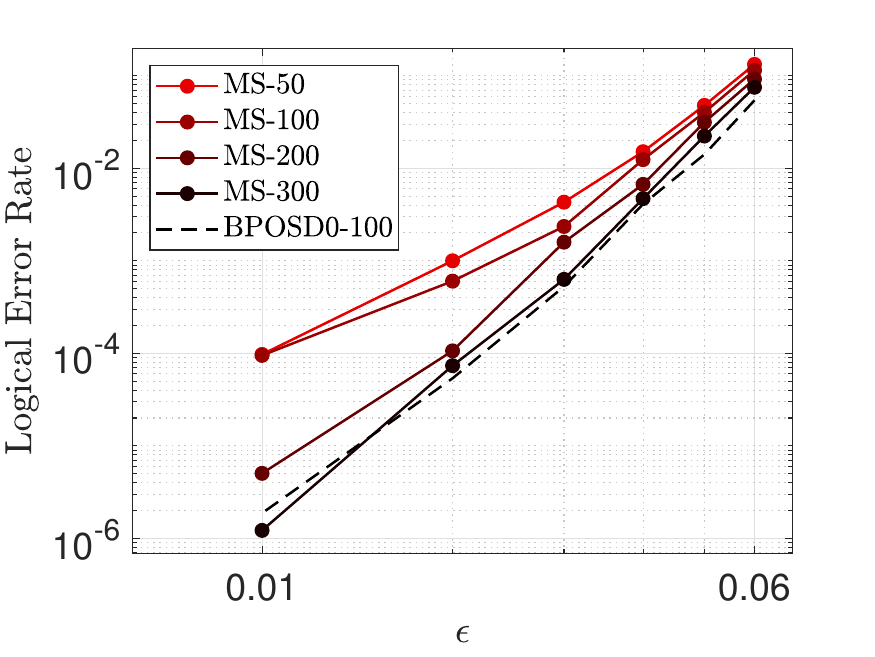}
    \caption{ Performance of MS decoding compared with BPOSD0 for a length 240 quantum Margulis code, for different amounts of maximum decoder iterations.}
    \label{fig:nms-osd_comparison1}
\end{figure}

\begin{figure}
    \centering 
    \includegraphics[width=0.5\textwidth]{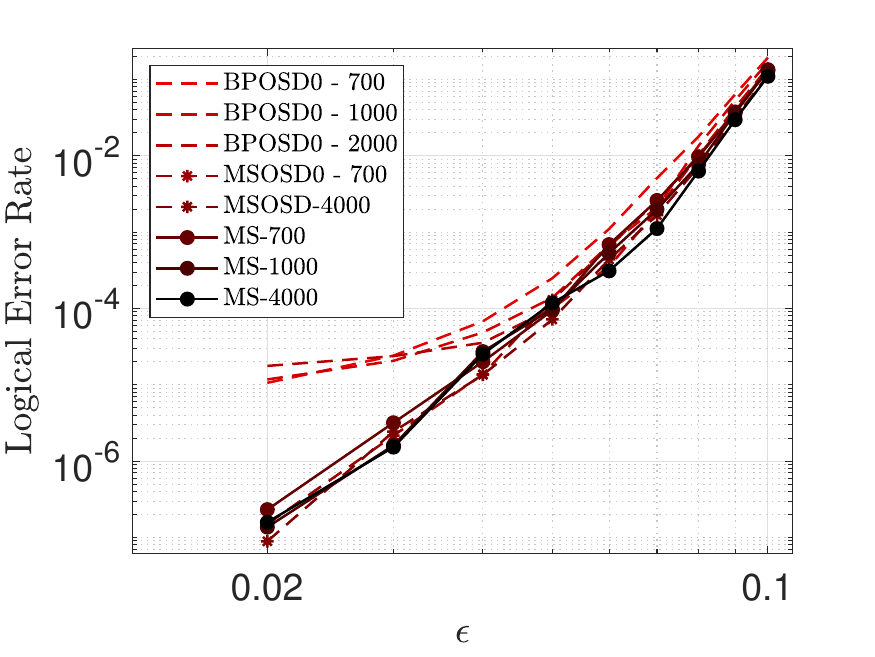}
    \caption{Performance of MS decoding compared with BPOSD0 and MSOSD0 for a length 642 quantum Margulis code, for increasing number of maximum iterations.}
    \label{fig:nms-osd_comparison2}
\end{figure}

A more involved approach is to think of the decoder as a non-linear dynamical system, and evaluate its \textit{Bethe free-entropy} \cite{declerq_faid_itw}. The Bethe free-entropy is a function that has its local minima corresponding to the convergence to a codeword or stabilizers, and its global minimum is reached when all the messages take the maximum reliability. By plotting the Bethe free-entropy of the decoder at the $k$-th iteration, as a function of the one at the $(k-1)$-th iteration, the oscillatory behavior of the decoder can be observed by the entropy revolving, while the convergence of the decoder makes the entropy decrease monotonically. To compute the Bethe free-entropy of the nMS decoder, we first transform the messages from the log-likelihood domain to the probability domain. Let $m_{j\rightarrow i}(0)$ denote the probability of the bit $b_i$ being zero as evaluated by its neighboring check node $c_j$, and $m_{j\rightarrow i}(1)$ the probability of the bit $b_i$ being one.
If the check-to-variable message is denoted by $\mu_{j\rightarrow i}$, then 
\begin{equation}
   m_{j\rightarrow i}(0) = \frac{e^{\mu_{j\rightarrow i}}}{1+e^{\mu_{j\rightarrow i}}},
\end{equation}
and, obviously, $m_{j\rightarrow i}(1) = 1-m_{j\rightarrow i}(0)$. The free entropy $E_i$ corresponding to a variable node $v_i$ can be calculated as
\begin{equation}
    E_i = \log \left \{\sum_{u=0}^1 \prod_{j=1}^{d_v} m_{j\rightarrow i}(u) \right \}.
\end{equation}
Conversely, the free-entropy $E_j$ corresponding to each check node $c_j$ can be calculated as
\begin{equation}
    E_j = \log \left \{ \sum_{i_1+...+i_{d_c}=0} \prod _{j=1}^{d_c } m_{i_n\rightarrow j}(i_n) \right \}.
\end{equation}
Finally, the free-entropy on an edge between $v_i$ and $c_j$ $E_{ij}$ can be calculated as
\begin{equation}
    E_{ij} = \log \left \{  \sum_{u=0}^1  m_{j\rightarrow i}(u)  m_{i\rightarrow j}(u)\right \},
\end{equation}
and the Bethe free-entropy at the $k$-th iteration is defined as 
\begin{equation}
    E^{(k)} = \sum_{i=1}^n E_i + \sum_{j=1}^m E_j - \sum_{(i,j) \in \mathcal{T}}^n E_{ij}.
    \label{eq:entropy}
\end{equation}
 In Fig. \ref{fig:bethe} we illustrate the Bethe free-entropy of nMS while decoding the error described before in a BB code and in the quantum Margulis code. In the figure, each value is plotted as a function of its previous value; this is to highlight the oscillatory behavior. For the BB code, the decoder is periodically attracted to a local minimum, but never converges, while for the quantum Margulis code it is able to reach its global minimum. In both Figs. \ref{fig:weight-iter}, \ref{fig:bethe} is also evident how the decoder needs to increase its entropy (which is correlated with $W$) before being able to converge.
We validate our observation through simulations, illustrated in Fig. \ref{fig:nms-osd_comparison1}-\ref{fig:nms-osd_comparison2}, where we compare the performance of a nMS and OSD decoders, for different number of maximum iterations, for quantum Margulis codes of length 240 and 642. In Fig. \ref{fig:nms-osd_comparison1}, we compare the nMS performance for increasing iterations with the  OSD after nMS (MSOSD) on a length 240 quantum Margulis code; it is clear how, by increasing the amount of iterations, the decoder is able to correct more and more errors, eventually approaching the OSD decoder. This suggests that the behavior shown in Fig. \ref{fig:bethe} takes much more iterations for some error patterns.
A similar phenomenon happens in  Fig. \ref{fig:nms-osd_comparison2}, where we also compare the performance of the BP-OSD, on a length 642 quantum Margulis code. It is evident how the nMS decoder is superior to the standard BP decoder, at a point where even the OSD post-processing is not able to decrease the error floor. This is the main reason behind our focus on nMS decoder, besides the fact that the nMS has also lower complexity, as the check node update rule is much faster in hardware. We notice that, in both cases, the maximum number of iterations required to approach the OSD decoder is similar, or slightly longer, than the code's blocklength.

\subsection{Further investigations}
To further investigate the role of graph asymmetry, we proceeded with the two following experiments.
First, we tried to construct other 2BGA codes based on non-Abelian groups (for instance, we utilized the dihedral group $D_n$); if a 2BGA code based on non-Abelian group does not show this decoding behavior, we can rule out this hypothesis. However we couldn't design a code whose parameters (length, degrees, rate and girth) would match the ones of the quantum Margulis codes. Indeed, it is highly non-trivial to find a group and a set of generators able to construct a 2BGA code with positive rate and also girth $g>4$ (at least for blocklengths comparable to our quantum Margulis codes). In this sense, our codes are quite unique.

In the second experiment, we analyzed several length-240 quantum Margulis codes and the length-288 BB code from \cite{bravyi2024high} with respect to graph diameter $\delta$, average path length $\bar{\delta}$, and spectral gap $\lambda$ (both families have girth $g=6$) We choose to compare with the length-288 BB code because it has similar blocklength and girth, but it is constructed from an Abelian group (the cyclic group $C_n$).

Graph diameter is directly related to message-passing performance \cite{tanner_group_ldpc}. The diameter of a graph is the maximum over all minimum distances between pairs of vertices. Since messages become dependent after $\lfloor g/2\rfloor$ iterations, if $\delta \gg g$ then remote vertices will necessarily receive statistically dependent information. Ideally, one seeks codes with large girth but small diameter. For all length-240 quantum Margulis codes we computed $\delta = 9$, whereas the BB code has $\delta = 8$. The average path length is $\bar{\delta}\approx 4.8$ for the quantum Margulis code and $\bar{\delta}=5$ for the BB code.

The spectral gap $\lambda$ is the difference between the moduli of the two largest eigenvalues of the adjacency matrix, and it directly measures edge expansion. A graph with a large spectral gap has high vertex connectivity, which leads to faster mixing of messages in iterative decoding (this is again connected to small diameter in expander graphs). For the quantum Margulis codes we found $\lambda \approx 0.3$, while the BB code has $\lambda = 0.7$. Thus, in these specific instances, the BB code exhibits significantly better expansion, consistent with its smaller diameter.

In summary, although the quantum Margulis code and the BB code share the same girth, the former has a larger diameter, roughly the same average path length, and a smaller spectral gap. Nevertheless, Min-Sum decoding successfully converges in the quantum Margulis code even when the stabilizers are symmetric, whereas this does not occur for BB codes (see Section \ref{sec:performance}). While this is not a definitive proof, it provides strong evidence that graph asymmetry is indeed responsible for this behavior. Even though there may be other non-Abelian codes with this feature, only the quantum Margulis code is able to achieve such large girths within short blocklengths, and positive dimension, making it one of the strongest families in his class.

In conclusion, quantum Margulis codes offer a significant advantage over other QLDPC codes by enabling decoding with a standard Min-Sum (MS) decoder, eliminating the need for OSD post-processing. The OSD step, which relies on Gaussian elimination of the parity-check matrix, scales as $\mathcal{O}(n^3)$, making it computationally expensive. This exponential complexity prevents low-latency decoding for moderate-to-large blocklengths, making OSD unsuitable for scalable fault-tolerant quantum systems.

\section{Performance of quantum Margulis codes}
\label{sec:performance}
\begin{figure}
    \centering
    \includegraphics[width=.5\textwidth]{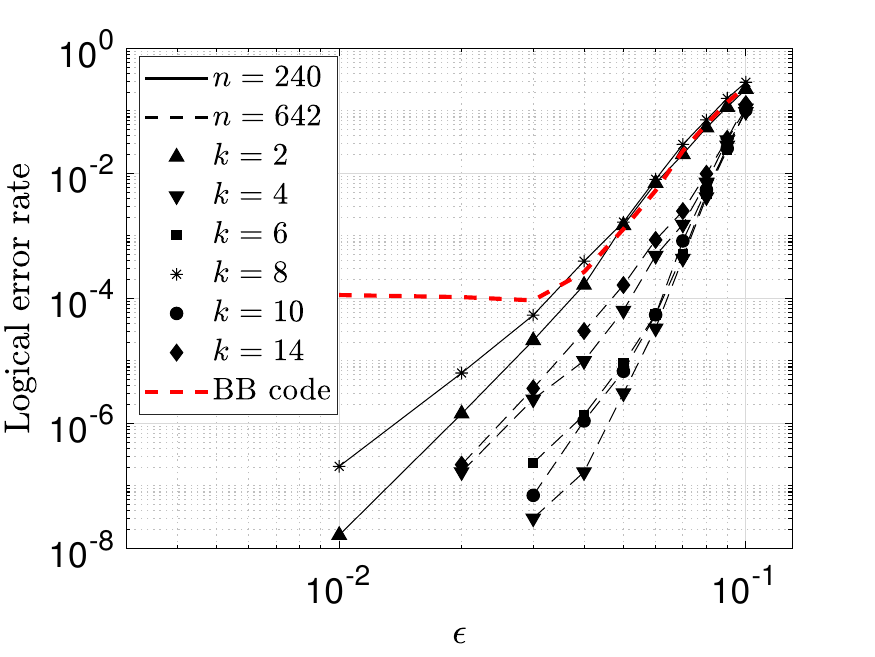}
    \caption{Simulation results for several codes sampled from the output of Algorithm \ref{alg:obtain_generators}.}
    \label{fig:simulations}
\end{figure}

\begin{figure}
    \centering 
    \includegraphics[width=0.5\textwidth]{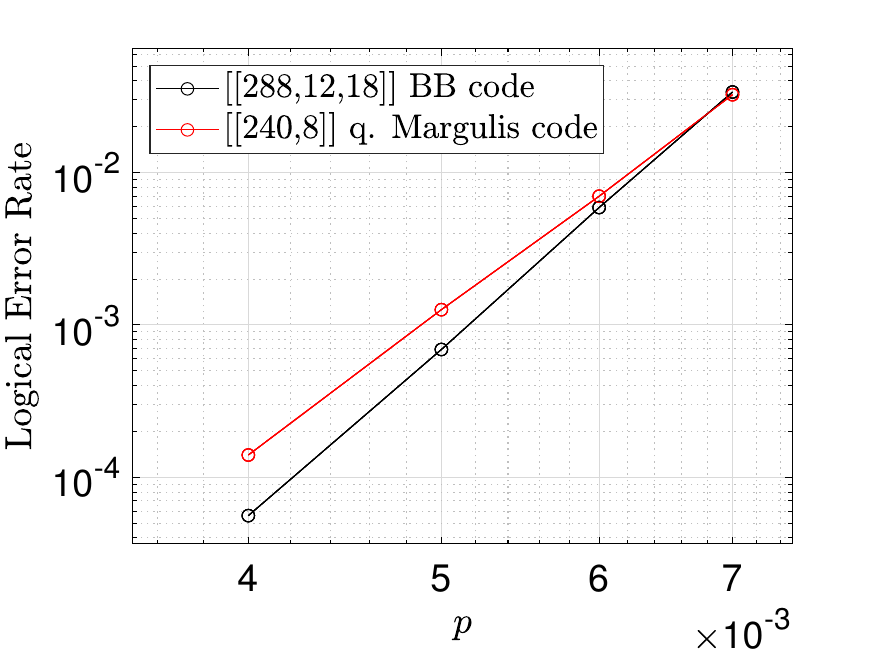}
    \caption{Performance comparison of a BB code and a quantum Margulis code under circuit-level noise. Each measurement is repeated four times, and decoding is performed using MSOSD10 on the circuit-level Tanner graph.}
    \label{fig:circuit-level}
\end{figure}

In this Section, we run extensive Monte Carlo simulations to validate the performance of quantum Margulis codes. 
We use Algorithm \ref{alg:obtain_generators} to generate several codes of length 240 and 642 (corresponding to $p=5,7$ respectively) and different dimension\footnote{All the codes constructed in this paper are available on GitHub \cite{git_repo}.}. For each code, we simulate nMS decoding under code capacity noise, where $\epsilon$ denotes the depolarizing probability. The nMS decoder utilizes a flooding (parallel) schedule, and a normalization factor $\beta = 0.875$ for all codes. We run a maximum of 300 iterations for $n=240$, and of 700 iterations for $n=642$. For each value of depolarizing probability $\epsilon$, the simulation is carried out such that $1\cdot 10^5$ samples and 20 decoding failures are collected. All the codes with $n=240$ have girth 6, while the ones with $n=642$ have girth 8. All the codes are $(3,6)$ regular.
We have included the codes that showed the best decoding performance in Fig. \ref{fig:simulations}, where the codes with $n=240$ are represented by solid lines, and the codes with $n=642$ are represented by dashed lines. Different values of $k$ are represented by the markers depicted in the legend. We also included simulation results for the $\llbracket 288,12,18 \rrbracket$ BB code from \cite{bravyi2024high} (in red), decoded with nMS, using $\beta = 0.875$ and for a maximum of 300 iterations. Such code exhibits a very strong error floor mostly due to uncorrectable errors in symmetric stabilizers. Among the codes with $n=240$, the one with lower error rate has $k=2$, able to approach approximately a logical error rate of $10^{-8}$ for $\epsilon = 0.01$.  Among the codes with $n=642$, the one with highest $k=14$ is also the one with weaker performance, approaching a logical error rate of approximately $2\cdot 10^{-6}$ for $\epsilon = 0.02$, while the best is the one with $k=2$, able to approach a logical error rate of $3\cdot 10^{-8}$ for $\epsilon = 0.03$. It is worth noticing the code with $k=10$, that approaches a logical error rate of around $7\cdot 10^{-8}$ for $\epsilon = 0.03$, which is remarkable given its higher rate. 
Finally, in Fig. \ref{fig:circuit-level} we compare the performance of the $\llbracket 240,8 \rrbracket $ quantum Margulis code and the $\llbracket 288,12,18 \rrbracket $ BB code under circuit-level noise. We use the $\mathtt{stim}$ package \cite{gidney2021stim} to sample errors from the circuit. The syndrome extraction is repeated 4 times, and the decoding is performed using MSOSD10 on the circuit-level Tanner graph, for a maximum of 1000 iterations. For this specific case, we utilize the BPOSD implementation available in the $\mathtt{ldpc}$ Python package \cite{roffe_decoding_2020}. Each value is obtained after simulating $10^6$ error samples. From the Figure, it is evident how, in this setting, the quantum Margulis code has slightly worse performance than the BB code, despite the superior performance under depolarizing noise.
In the circuit-level Tanner graph, each possible circuit fault is represented as an additional variable node (for more details see \cite{higgott_improved_2023, gong2024lowlatencyiterativedecodingqldpc}). Decoding must therefore be carried out on an enlarged Tanner graph that contains many short cycles and highly irregular node degrees. Under these conditions, standard Min-Sum decoding is not viable, and BPOSD (or other circuit-level decoders, typically more complex than BP) becomes necessary. In this regime, the advantage provided by the asymmetry of the original Tanner graph is effectively lost, while the structural complications of the circuit-level graph become the main limitation.
For these reasons, it is difficult to identify the precise cause of the poorer performance of the quantum Margulis code relative to the BB code under circuit-level noise. Improved MSOSD hyperparameters or a different stabilizer-measurement schedule might help, although a BB-like schedule can already be implemented without difficulty. Another possibility is that the quantum Margulis code simply has a smaller minimum distance, since better performance under the depolarizing channel does not guarantee a larger minimum distance.
A detailed investigation of this behavior lies beyond the scope of the present work. We include the comparison mainly for completeness and to underscore the need for circuit-level decoders capable of exploiting the structure of the original code. We also note that Monte Carlo simulations of multiple rounds of repeated measurements at this blocklength, decoded with BPOSD, are computationally expensive, which limits the extent of our study.

\section{Conclusions}
\label{sec:conclusions}
In this work, we introduced quantum Margulis codes, a new class of QLDPC codes derived from Margulis’ classical LDPC construction through the 2BGA framework. Our construction algorithm allows for control over the girth of the Tanner graph, ensuring that short cycles are minimized. Using this approach, we generated codes with girths of at least six and eight, equipped with an inherent asymmetry that aids the standard nMS decoder to converge without the need of the computationally expensive OSD, when considering code capacity noise. This represents a significant improvement in decoding complexity, reducing it from $\mathcal{O}(n^3)$ to $\mathcal{O}(n)$, making these codes more viable for practical quantum error correction. Future directions may include a more comprehensive study of 2BGA codes obtained from different groups, to verify their performance under different decoders. Also, it is possible that the particular structure of some groups translate into good properties of the corresponding 2BGA codes; for instance, groups like $PSL(2,q)$, with $q$ prime may originate codes with good expansion properties, that could be characterized by their error correction capability. Other groups may allow larger graph automorphisms, or present geometrical structures that could be more suitable for certain hardware implementations.
In any case, when it comes to decoding circuit-level noise, the decoding advantage of quantum Margulis codes is lost, as we are forced to use OSD post-processing due to the different structure of the decoding matrix. This highlights the urgency of developing novel decoders for circuit-level noise, that are able to exploit the optimized structure of the original Tanner graph while accounting for noise correlations, avoiding the need of OSD.



\input{references}
\end{document}

%% file: figures/square.tikz
\begin{tikzpicture}[scale=2]

    \tikzset{arrow/.style={-{Stealth[length=3mm,width=2mm]}, line width=0.5pt}}
    \draw[arrow] (0,0) -- (1,0);
    \draw[arrow] (1,0) -- (1,1);
    \draw[arrow] (0,1) -- (1,1);
    \draw[arrow] (0,0) -- (0,1);
    
    \node[circle, fill=black, inner sep=1pt, label={[shift={(-0.7,-0.5)}]$(g,00)$}] (A) at (0,0) {};
    \node[circle, fill=black, inner sep=1pt, label={[shift={(0.7,-0.5)}]$(ga,10)$}] (B) at (1,0) {};
    \node[circle, fill=black, inner sep=1pt, label={[shift={(0.8,0.01)}]$(bga,11)$}] (C) at (1,1) {};
    \node[circle, fill=black, inner sep=1pt, label={[shift={(-0.7,0.01)}]$(bg,01)$}] (D) at (0,1) {};
    
    \node[anchor=north] at (0.5,0) {$a$};
    \node[anchor=west] at (1,0.5) {$b$};
    \node[anchor=south] at (0.5,1) {$a$};
    \node[anchor=east] at (0,0.5) {$b$};
    
\end{tikzpicture}

%% file: figures/square2.tikz
\begin{tikzpicture}[scale=2]

    \tikzset{arrow/.style={-{Stealth[length=3mm,width=2mm]}, line width=0.5pt}}
    \draw[arrow] (1,0) -- (0,0);
    \draw[arrow] (1,1) -- (1,0);
    \draw[arrow] (1,1) -- (0,1);
    \draw[arrow] (0,1) -- (0,0);
    
    \node[circle, fill=black, inner sep=1pt, label={[shift={(-0.7,-0.5)}]$(g,00)$}] (A) at (0,0) {};
    \node[circle, fill=black, inner sep=1pt, label={[shift={(0.7,-0.5)}]$(ga,10)$}] (B) at (1,0) {};
    \node[circle, fill=black, inner sep=1pt, label={[shift={(0.8,0.01)}]$(bga,11)$}] (C) at (1,1) {};
    \node[circle, fill=black, inner sep=1pt, label={[shift={(-0.7,0.01)}]$(bg,01)$}] (D) at (0,1) {};
    
    \node[anchor=north] at (0.5,0) {$a^{-1}$};
    \node[anchor=west] at (1,0.5) {$b^{-1}$};
    \node[anchor=south] at (0.5,1) {$a^{-1}$};
    \node[anchor=east] at (0,0.5) {$b^{-1}$};
    
\end{tikzpicture}

%% file: Algorithms/get_marg_gen.tex
\State $S\gets d_c$ random elements of $G$ 
\State $\mathcal{C} \gets \emptyset$ \Comment{Cycle to remove}
\State $ f \gets 1 $, $\ell \gets 1 $
\While{$f = 1$ and $\ell < \gamma /2$}
    \If{$\ell > 1$}
        \State Replace $s \in \setc$ with $s' \in G\setminus S$ in $S$.
    \EndIf
    \State $ S,f, \mathcal{C}, \ell \gets \mathtt{generate\_tree}(S, \gamma)$
\EndWhile
\If{$G$ is non-Abelian}
    \For{$g \in G$}
        \State $S' \gets gA \cup Bg$
        \State $f  \gets \mathtt{generate\_tree}(S',\gamma)$
        \If {$f= 1$}
            \State Restart from line 1.
        \EndIf
    \EndFor
\EndIf

%% file: tikz/tree1.tex
    \begin{tikzpicture}[,
        level distance=1cm,
        sibling distance=2.5cm,
        every node/.style={draw},
        level 1/.style={sibling distance=2.5cm},
        level 2/.style={sibling distance=1cm},
        check/.style={edge from parent/.style={draw,latex-}},
        var/.style={edge from parent/.style={draw,-latex}}
    ]
        \node[rectangle, fill=red] (r) {} 
            child[var] {node[circle, fill=black!90] {}  
                child[check] {node[rectangle, fill=red] {}}
            }
            child[var] {node[circle, fill=black!90] (a3) {} 
                child[check, xshift=1.2cm] {node[rectangle, fill=red] (a1) {}}
            }
            child[var] {node[circle, fill=black!10] (a2) {} 
            }
            child[var] {node[circle, fill=black!10] {} 
                child[check] {node[rectangle, fill=red] {}}
            };
            \draw[-latex,thick,red] (a1) -- (a2);
            \draw[-latex,thick,red] (a1) -- (a3);
            \draw[-latex,thick,red] (r) -- (a2);
            \draw[-latex,thick,red] (r) -- (a3);
    \end{tikzpicture}

%% file: tikz/tree2.tex
    \begin{tikzpicture}[scale=0.20,
        level distance=3.5cm, 
        sibling distance=1cm, 
        every node/.style={draw}, 
        level 1/.style={sibling distance=10cm}, 
        level 2/.style={sibling distance=5cm}, 
        level 3/.style={sibling distance=3cm}, 
        check/.style={edge from parent/.style={draw,latex-}},
        var/.style={edge from parent/.style={draw,-latex}}
    ]
        \node[rectangle, fill=red] (r) {} 
            child[var] {node[circle, fill=black!90] {}  
                child[check] {node[rectangle, fill=red] {}
                    child[var] {node[circle, fill=black!90] {}}
                    child[var] {node[circle, fill=black!10] {}}
                    child[var] {node[circle, fill=black!10] {}}
                }
            }
            child {node[circle, fill=black!90] (a1) {} 
                child {node[rectangle, fill=red] (a2) {}
                    child[var] {node[circle, fill=black!90] {}}
                    child[var] {node[circle, fill=black!10] {}}
                    child[xshift=1.9cm] {node[circle, fill=black!10] (a3) {}}
                }
            }
            child {node[circle, fill=black!10] (b1) {} 
                child {node[rectangle, fill=red] (b2) {}
                    child[var] {node[circle, fill=black!90] {}}
                    child[var] {node[circle, fill=black!90] {}}
                }
            }
            child[var] {node[circle, fill=black!10] {} 
                child[check] {node[rectangle, fill=red] {}
                    child[var] {node[circle, fill=black!90] {}}
                    child[var] {node[circle, fill=black!90] {}}
                    child[var] {node[circle, fill=black!10] {}}
                }
            };

            \draw[-latex,thick,red] (a2) -- (a1);
            \draw[-latex,thick,red] (r) -- (a1);
            \draw[-latex,thick,red] (a2) -- (a3);
            \draw[-latex,thick,red] (r) -- (b1);
            \draw[-latex,thick,red] (b2) -- (b1);
            \draw[-latex,thick,red] (b2) -- (a3);

    \end{tikzpicture}

%% file: Algorithms/generate_tree.tex
\begin{algorithmic}[1]
        \State $\mathcal{L}_{\ell+1} \gets \emptyset$
        \For{$l \in \mathcal{L}$}
            \If{$\ell$ is even}
                \State $\mathcal{L}_{\ell+1} \gets l \cdot a_j\ \forall a \in A \setminus a_l$ 
                \State $\mathcal{L}_{\ell+1} \gets b_j \cdot l\ \forall b \in B \setminus b_l$
            \Else
                \State $\mathcal{L}^R, \mathcal{L}^L \gets \emptyset$
                \State $\mathcal{L}^R \gets l \cdot a_j^{-1}\ \forall a_j^{-1} \in A^{-1} \setminus a_l^{-1}$
                \State $\mathcal{L}^L \gets b_j^{-1} \cdot l\ \forall b_j^{-1} \in B^{-1} \setminus b_l^{-1}$
                \State $\mathcal{L}_{\ell+1} \gets \mathcal{L}^L \cup \mathcal{L}^R$
            \EndIf
        \EndFor
        \If{ $\ell = \gamma/2$}
            \State \Return 
        \EndIf
        \If{No vertex with more than one incident edge in $\mathcal{L}_{\ell+1}$}
            \State $\mathtt{generate\_tree}(S, \mathcal{L}_{\ell+1}, \gamma, \ell+1)$
        \Else
            \State $\setc \gets $ cycle edges
            \State $f \gets 1$
        \EndIf
        \State \Return
    \end{algorithmic}

%% file: tikz/symmstab.tikz
\begin{tikzpicture}[scale=0.8, every node/.style={transform shape}]

\tikzstyle{circle_node} = [circle, draw, fill=black!10, inner sep=0pt, minimum size=3mm]
\tikzstyle{square_node} = [rectangle, draw, fill=red, inner sep=0pt, minimum size=3mm]
\tikzstyle{circle_alt_node} = [circle, draw, fill=black!90, inner sep=0pt, minimum size=3mm]

\node[circle_alt_node] (a1) at (0, 0) {};
\node[square_node] (c1) at (1, 0) {};
\node[circle_node] (b1) at (2, 0) {};
\node[square_node] (c2) at (2.8, -1) {};
\node[square_node] (c3) at (2.8, -3) {};
\node[circle_alt_node] (a2) at (3.5, -2) {};
\node[circle_node] (b2) at (2, -4) {};
\node[square_node] (c4) at (1, -4) {};
\node[circle_alt_node] (a3) at (0, -4) {};
\node[square_node] (c5) at (-0.8, -1) {};
\node[square_node] (c6) at (-0.8, -3) {};
\node[circle_node] (b3) at (-1.5, -2) {};

\draw (a1)--(c1)--(b1)--(c2)--(a2)--(c3)--(b2)--(c4)--(a3)--(c6)--(b3)--(c5)--(a1);
\draw (a1)--(b2) node[pos=0.8,square_node]{};
\draw (b1)--(a3) node[pos=0.2,square_node]{};
\draw (a2)--(b3) node[pos=0.8,square_node]{};

\end{tikzpicture}